\documentclass[%
 reprint,
 amsmath,amssymb,
 aps,
]{revtex4-1}

\usepackage{graphicx}
\usepackage{dcolumn}
\usepackage{bm}

\begin{document}

\preprint{APS/123-QED}

\title{Experimental Evidences Supporting the Extension of the Equivalence Principle to Electromagnetic Fields \\}

\author{Clovis Jacinto de Matos}%
\email{clovis.de.matos@esa.int}
\email{clovis.de$_$matos@mailbox.tu-dresden.de}
\affiliation{
European Space Agency, Headquarters, 8-10 rue Mario Nikis, 75015 Paris, France \\ and Institute of Aerospace Engineering, Technische Universit\"{a}t Dresden, Marschnerstrasse 32, 01307 Dresden, Germany }%

\author{Murat \"{O}zer}
\email{mhozer@yildiz.edu.tr}
\affiliation{%
Department of Physics, Faculty of Arts and Sciences, 
Y{\i}ld{\i}z Technical University, 34220 Esenler, Istanbul, Turkey
}%

\author{Grzegorz Lukasz Izworski}%
\email{Grzegorz.Lukasz.Izworski@esa.int}
\affiliation{%
 European Space Agency, ESTEC,  Keplerlaan 1, 2201 AZ Noordwijk, the Netherlands
\\}%
\date{\today}

\begin{abstract}
  The principle of equivalence postulating that an acceleration is indistinguishable from gravity by any experiment, is valid within families of particles having the same passive gravitational to 
inertial mass ratio $m_p/m_i$. Presently experimental observations indicate that we live in a universe with one single family for which $m_p/m_i=1$, but if we consider the imaginary case of a 
universe with several particle families having different $m_p/m_i$, the principle of equivalence would still apply to each one of them. On the basis of this generalized formulation of the 
equivalence principle, which becomes relative to sets of particles, and that we designate as the \textit{single-particle equivalence principle}, one demonstrates that inertial frames can also be 
implemented for sets of electrically charged particles, with the same charge-to-mass ratio $q/m_i$, accelerating in homogeneous electric and / or magnetic fields (by analogy with the case of 
particles in free fall in a homogeneous gravitational field). Experimental evidences in support of the proposed extension of the principle of equivalence to electric fields are presented. These 
consist in the Witteborn-Fairbank experiment which revealed that electrons do not fall in the Earth gravitational field, and the well know fact that electric charges do not radiate either when in 
free fall in a homogeneous gravitational field, or when being accelerated by a homogeneous electric field. The London moment in rotating superconductors is also supporting the proposed extension 
of the equivalence principle. Prospects for the future exploration of the consequences of the proposed theoretical scheme to unify electromagnetism with gravitation in the paradigm of curved 
space-time, are also briefly suggested.

\begin{description}
\item[Key words]
general relativity, principle of equivalence, gravitation, electromagnetism 

\item[PACS numbers]
04.50.+h, 04.20.Cv. 41.2.Cv
\end{description}
\end{abstract}

\maketitle

\section{\label{Intro} Introduction}

The most fundamental experimental observation about the physical nature of gravitation, is that any bodies subjected only to a homogeneous gravitational field, have identical motions, independent 
of their masses, for identical initial conditions of position and velocity. Newton mechanics and Einstein theory of general relativity correctly accounted for this observation in terms of the exact
equality between the inertial mass $m_i$, and the passive gravitational mass $m_p$, of any physical body, $m_i=m_p$ (also designated as the weak equivalence principle). However these two types of 
masses are very different. On the one side, the inertial mass of a body is responsible for the resistance it offers to forces that can change its state of motion, on the other side the passive 
gravitational mass of a body is its mass that is acted upon by the gravitational field at its location. The cause of the observed equality between these two different masses is presently an active 
field of research in physics \cite{Gine}. The authors would like to stress that the present paper does not aim to offer an explanation for this well proven experimental fact \cite{Baessler}. 

Einstein made the fundamental observation that the motion of any bodies with respect to a non-inertial reference frame, moving with constant acceleration away from any gravitational fields, have 
identical motions, independent of their respective inertial masses, for identical initial conditions. On this basis he postulated the \textit{principle of equivalence}, formulating the equivalence 
between non-inertial reference frames moving with constant accelerations and homogeneous gravitational fields. This allowed to understand, in the context of the theory of general relativity, that 
the motion of bodies in a gravitational field is merely inertial motion in curved spacetime, i.e. motion in the absence of any forces acting on the body. Let us emphasize that the principle of 
equivalence does not require \textit{per se} the equality between the inertial and the gravitational mass of physical bodies. Because the fact that the trajectories of test particles, with respect 
to an accelerated reference frame (in the absence of any gravitational fields), are independent of their respective inertial masses, is a pure cinematic effect, which has nothing to do with the 
equality between inertial and gravitational masses. This is also designated as the strong version of the equivalence principle, since it contrasts in this respect with the weak equivalence 
principle.  

How could we formulate the principle of equivalence in an imaginary world where the equality between inertial and gravitational mass is not verified for all bodies? Following the work of 
\"{O}zer \cite{Ozer_2000}, we show in section \ref{Section_1} how the answer to this question allows to extrapolate the principle of equivalence to the case of electrically charged matter moving 
under the influence of electric fields. In section \ref{section_2} we present the experimental evidence, already existing, in support of the extension of the principle of equivalence to the case 
of electric fields, by first showing how it can account for the Witteborn-Fairbank experiment, which reported the absence of free fall for electrons in the Earth gravitational field, and in a 
second step by showing how it is consistent with the experimental fact that electric charges in free fall in a constant gravitational field do not radiate electromagnetic waves. Afterwards we 
show that the extension of the principle of equivalence to magnetic fields is supported by the magnetic London moment, exhibited by superconductors when set into rotation. We conclude by 
highlighting the phenomenological framework in support of the inertial bridge between gravitation and electromagnetism, and with some proposals on possible future experimental work, to further 
investigate the physical consequences of the proposed extension of the principle of equivalence to electromagnetic fields.  

\section{\label{Section_1} General Principle of Equivalence for Gravitation and Electromagnetic Fields}

According to Einstein's principle of equivalence, a reference frame in free fall under the single influence of a homogeneous gravitational field is locally equivalent to an inertial frame, i.e 
a frame which is at rest or moves with constant velocity with respect to the gravitational source, in other words, a frame on which are applied no forces at all. The relative acceleration 
between test masses in the falling frame and the falling frame is null, thus particles will exhibit weightlessness in the falling frame. By a simple coordinate change the gravitational force 
disappears, and gravity becomes a pure geometric phenomena as fully developed in the theory of general relativity. The space-time metric in the falling frame will be of Minkowski type over the 
entire volume delimited by the falling frame boundaries, and will be independent of the test masses included in the falling frame, and of the mass of the falling frame itself, assuming that a 
reference frame needs to be physically implemented by a physical body having inertial and gravitational mass as well. The independence of geometry with respect to test bodies, and reference frame 
masses is due to the fact that all particles have the same gravitational to inertial mass ratio. If this would not be the case, what would it mean for space-time geometry? Before we answer, it is 
to be noted that although space-time metric does not depend explicitly on the test bodies masses, when their inertial and gravitational masses are equal, it can only be measured by observing the 
motion of test bodies. Therefore spacetime geometry only acquires, locally, an operational meaning through the local presence of test bodies.      

\subsection{\label{section_1_1}Inertia in Gravitational Fields}

Let us consider an imaginary world, where two bodies $A$ and $B$, with different gravitational to inertial mass ratios $m_p/m_{i}$ and $M_p/M_{i}$ respectively, are in free fall under the single 
influence of an homogeneous gravitational field $g$. Assuming identical initial conditions, the relative acceleration between the two bodies will be:
\begin{equation}
a_{rel}=\Big(\frac{m_{p}}{m_{i}}-\frac{M_{p}}{M_{i}}\Big) g. \label{e1}
\end{equation}
If the body $A$ is the reference frame and the body $B$ is the test mass (arbitrary choice), the relative acceleration will only be null if the test particle has the same gravitational to mass 
ratio as the particle defining the origin of the freely falling reference frame. We conclude that a freely falling reference frame attached with a particle having a given $m_p/m_i$ ratio is only 
an inertial frame for test particles having identical gravitational to inertial mass ratio $m_p/m_i$. In other words, in this imaginary world, sets of particles sharing the same $m_p/m_i$ ratio, 
would form different local inertial frames. Each set would travel in its own space-time geometry following their own geodesics, side by side, with respect to the same gravitational source. 
Although we neglect the effect of the test masses on the source of the gravitational field, they would play an active role in defining locally the components of the metric tensor. This could be 
consistent, to the extent that only one test particle can occupy one space-time point at a time. We are thus led with \"{O}zer \cite{Ozer_2000} to state the \textit{"single-particle equivalence 
principle: It is impossible to distinguish the fictitious inertial forces from the real gravitational forces in a local region containing a single particle or set of particles with the same 
$m_p/m_i$ ratio."} Space-time curvature at a given spacetime point would thus result from the combined effects of a gravitational field source and the inertial to gravitational mass ratio of the 
test bodies present at this specific point. In this imaginary world, two test bodies with different inertial to gravitational mass ratios, evolving near the same gravitational source would 
experience different curvature at the same space-time point, which they would occupy one after the other. In this theoretical framework Einstein field equations  adopt the following form:
\begin{equation}
R^{\mu \nu}-\frac{1}{2} g^{\mu\nu} R=\frac{8\pi G}{c^4}\frac{m_p}{m_i}T^{\mu\nu}, \label{e2}
\end{equation}
where $c$ is the speed of light in vacuum, $G$ is Newton's gravitational constant,  $T^{\mu\nu}$ is the energy momentum tensor of the gravitational source, with Greek indices $\mu=0,1,2,3$ 
designating the time and space components, $R^{\mu \nu}$ and $R$ are the Ricci curvature tensor and Ricci scalar respectively, $g^{\mu\nu}$ is the spacetime metric tensor, and ${m_p}/{m_i}$ is 
the gravitational to inertial mass ratio of the test particles used to measure locally the space-time curvature generated by the gravitational source and the test particle.

Therefore we are led to the conclusion that the geometric nature of gravitational fields, in terms of space-time curvature does not imperatively require the universal equality between inertial 
and gravitational masses of all physical bodies present in the universe (and does not require one single universal value for the ratio between inertial and gravitational mass). If this equality 
is verified, the inertial to gravitational mass ratio is one, and  it will be impossible to notice explicitly the effect of the test masses on the local value of space-time curvature, which will, 
in this particular case, only depend on the source of the gravitational fields (this is the current situation in the theory of general relativity).

\subsection{\label{IEF} Inertia in Electromagnetic Fields}

\subsubsection{\label{section_1_2}Inertia in Electric Fields}

It is straight forward to extend the analysis made in section \ref{section_1_1} to the case of electric fields, if we replace the homogeneous gravitational field by an homogeneous electric field, 
and we use particles with electric charge-to-mass ratio $q/m_i$ as test bodies, and to define reference frames. Let us consider two particles with electric charge-to-mass ration $q/m_i$ and $Q/M_i$
, falling freely under the single influence of a downward electric field $E$. The relative acceleration between the two particles is given by:
\begin{equation}
a_{rel}=\Big(\frac{q}{m_i}-\frac{Q}{M_i}\Big)E, \label{e3}
\end{equation}
and is null when the electric charge-to-mass ratios are equal, $q/m_i=Q/M_i$, for this case a reference frame attached to one of the particles can constitute a local inertial frame for the other. 
This indicates the possibility to achieve an inertial frame for a set of particles with identical charge-to-mass ratio being accelerated in a homogeneous electric field. 

We are therefore led to the \textit{electric analog of the single-particle equivalence principle} for the case of electrically charged matter: \textit{"It is impossible to distinguish the 
fictitious inertial force, due to a frame accelerating with constant acceleration $a=qE/m_i$, from the real electric force $qE$ (in the homogeneous electric field $E$), in a local region containing
 a single particle or particles with the same electric charge-to-mass ratio"}.

\subsubsection{\label{section_1_2_a} Inertia in Magnetic Fields}

It is possible to extend the arguments of section \ref{section_1_2} to the case of electrically charged particles, moving with constant velocity $v$ in a homogeneous magnetic field $B$. If we 
consider two particles with electric charge-to-mass ratio $q/m_i$ and $Q/M_i$ respectively, their relative acceleration, which in the present experimental situation is only caused by a homogeneous 
magnetic force, will be null if $q/m_i=Q/M_i$, i.e.
\begin{equation}
a_{rel}=\Big(\frac{q}{m_i}-\frac{Q}{M_i}\Big)vB. \label{f1}
\end{equation}
This indicates the possibility to realize an inertial frame for a set of particles with identical charge-to-mass ratio being accelerated in a homogeneous magnetic field. The corresponding 
formulation of the single-particle equivalence principle for electrically charged matter in homogeneous magnetic fields would be:  \textit{It is impossible to distinguish between the fictitious 
inertial force, due to a frame accelerating with constant Coriolis acceleration $a=2v\Omega=(q/m_i)vB$, when moving with velocity $v$ in a field of angular velocity $\Omega$ (the centrifugal 
acceleration is neglected with respect to the Coriolis acceleration, constraining the magnetic field to be weak), from the real magnetic force $qvB$ (in the homogeneous magnetic field $B$), in a 
local region containing a single particle or particles with the same electric charge-to-mass ratio.} 

\subsubsection{\label{FEE} Einstein Field Equations for Electromagnetism Revisited}

On the basis of the single-particle equivalence principle for homogeneous electric and magnetic fields, one can postulate the single-particle equivalence principle for electromagnetism: 
"\textit{All effects of a uniform electromagnetic field locally on a single particle or particles with the same electric charge-to-mass ratio are identical to the effects of a uniform acceleration 
of the reference frame}. From this principle one can understand electromagnetic fields in terms of space-time curvature, in straight analogy with the imaginary case of a universe where gravitational 
and inertial masses are different. Electrically charged bodies would curve space-time like masses do, and the $q/m_i$ ratios of test particles would also appear explicitly in the components of the 
space-time metric tensor together with the $m_p/m_i$ ratios of test masses. For the particular case of a universe where $m_p/m_i=1$ (which is the current experimental truth), only the $q/m_i$ would 
be left visible in the metric components for what concerns the influence of test bodies on space-time curvature, which can be calculated from the following field equations, for electric charges 
and currents: 
\begin{equation}
R^{\mu \nu}-\frac{1}{2} g^{\mu\nu} R=\frac{2 \mu_0}{c^2}\frac{q}{m_i}T_{CC}^{\mu\nu}, \label{e5}
\end{equation}
where $\mu_0$ is the magnetic permeability of vacuum, $q/{m_i}$ is the electric charge to mass ratio of the test particle used to measure locally the space-time curvature, generated by the source 
of electromagnetic fields and the test particle. $T_{CC}^{\mu \nu}$ is the charge current tensor, which has units of $A m^{-1} s^{-1}$, and corresponds to the time variation of the magnetic field 
strength generated by a distribution of electric current densities, not to be confused with the energy-momentum tensor for electromagnetic fields, which corresponds to electromagnetic energy 
densities, and is part of the standard Einstein-Maxwell equations \cite{Fuzfa}. Since this is not the main scope of the present paper, to  obtain additional information about the structure of the 
charge current tensor, $T_{CC}^{\mu \nu}$, and appreciate further the detailed formulation of the theory of general relativity on the basis of the single-particle equivalence principle for electric
 fields, the reader is invited to consult the reference \cite{Ozer_2000} and references therein. 

\section{\label{section_2} Experimental Evidences Supporting a Generalized Principal of Equivalence for Electromagnetic Fields}

The extension of the principle of equivalence to electromagnetic fields leads to three fundamental experimental predictions:

\begin{enumerate}
\item A set of electric charges, with the same charge-to-mass ratios, subject to a constant gravitational field, should not fall with respect to each other.
\item An electric charge in free fall under the single influence of a constant gravitational field, or being accelerated by a constant electric field, does not radiate electromagnetic waves.
\item A superconductor set into rotation is the source of a homogeneous magnetic field
\end{enumerate}

The first prediction has been observed by Witteborn and Fairbank but was wrongly interpreted by Schiff and Barnhill in terms of electrostatic interactions between different components of the 
experiment. The second prediction is related with how an electric charge submitted to a constant acceleration radiates or not. This question has been a wild source of controversy in theoretical 
as well as in experimental physics. In the present paper we argue that it does not due to the single-particle equivalence principle for electric fields. The magnetic field of rotating 
superconductors is known as the London Moment, and is routinely observed in the laboratory. It is associated with Cooper pairs supercurrents induced by the rotation of the superconductor bulk 
material. We show here how the London moment results directly from the absence of relative acceleration between Cooper pairs in the rotating frame.   

\subsection{\label{section_2_1}Witteborn-Fairbank Experiment on the Free Fall of Electrons}

Witteborn and Fairbank tested the equality between the inertial and gravitational mass of electrons, by measuring the time of flight of electrons falling freely in the Earth gravitational field 
in a shielded metallic drift tube. It was observed that the electron's were not falling, their free fall acceleration was null with an uncertainty of $\pm 0.09 g$ \cite{Witteborn}. This was 
interpreted by Schiff and Barnhill \cite{Schiff_1966} as being due to the action of gravity on the conduction electrons of the metal of the drift tube. They have shown that for falling charged 
test particles with mass M and Charge Q, the effective gravitational acceleration resulting from the effect of the electrons in the drift-tube, is given by\footnote{From now on we drop the 
subscript $i$ from the masses and all the masses are to be understood as inertial masses}:
\begin{equation}
g_{eff}=g\Big (1- \frac{m_e}{e} \frac{Q}{M} \Big), \label{e6}
\end{equation}
where $g$ is the earth gravitational acceleration, $m_e$ and $e$ are the mass and electric charge of the electrons in the metal of the drift tube. $g_{eff}=0$ when the charge-to-mass ratio of the 
falling test charge and of the electrons in the drift tube are equal, i.e when $Q=e$ and $M=m_e$, i.e when the charge-to-mass ratio of the particles in the drift tube walls are equal to the 
charge-to-mass ratio of the test particles falling inside the drift tube. 
	In the thought experiment of section \ref{section_1_2}, the acceleration $a$ caused by the constant electric field $E$, plays the role of the earth gravitational acceleration $g$ in 
Witteborn-Fairbank experiment.
\begin{equation}
a=\frac{q} {m} E. \label{e7}
\end{equation}
Substituting equation (\ref{e7}) in equation (\ref{e3}) one obtains the relative acceleration between the two accelerated charges in function of the acceleration caused by the applied constant 
electric field:
\begin{equation}
a_{rel}=a\Big(1-\frac{m} {q} \frac {Q}{M} \Big). \label{e8}
\end{equation}
The equation (\ref{e8}) for our thought experiment, and the equation (\ref{e6}) for Witteborn-Fairbank experiment are identical, when we consider the free fall of electrons, i.e when $Q=q=e$ and 
$M=m=m_e$. In other words when the charge-to-mass ratio of the two electric charges are identical, their relative acceleration is null in line with the single particle equivalence principle which 
we formulated in section \ref{section_1_2}. Therefore the Witteborn-Fairbank experiment is the experimental proof that the “single-particle principle of equivalence for electrically charged matter”
 is correct. A gravitational acceleration is indistinguishable from  an acceleration caused by an electric field , $a=q E/m = g$, therefore both must be interpreted, at their most fundamental level
, as space-time curvature, if the test-particle the observer and the reference frame (the drift tube walls) have all the same charge-to-mass ratio!

When establishing equation (\ref{e6}) Schiff and Barnhill have neglected the effect of the gravity on the drift tube ion lattice, which is positively charged, and should thus lead to an electric 
field, inside the drift tube, roughly ten thousand times larger than the one caused by the drift tube conduction electrons, and should have opposite direction (since the lattice ions have a mass 
ten thousand times higher than the electron and are positively charged). Thus Schiff and Barnhill physical explanation of Witteborn-Fairbank experiment, on the basis of the screening of the ionic 
lattice of the drift tube by conduction electrons is insufficient to account for the observed zero acceleration of falling electrons inside the drift tube, as recognized by Schiff and Barnhill 
themselves in a following paper \cite{Schiff_1970} (together with additional experimental data referenced therein). Here we argue that this is an additional experimental indication that one can 
only account properly for the witteborn-Fairbank experiment when the single-particle equivalence principle is extended to electrically charged matter, thus requiring that when we use electrons as 
test masses falling inside the drift tube one must also consider the action of gravity on the conduction electrons of the drift tube, if instead one uses ions as test masses, one should consider 
the action of gravity on the drift tube ion lattice \cite{note_1}. The vanishing quantity for the case of electrons in free fall in the drift tube is not the sum $\vec{E}+m_e \vec{g}/e$, 
equation (\ref{e7}), as initially claimed by Schiff and Barnhill, but rather $\Big(\frac{e}{m_e}-\frac{Q}{M}\Big)\vec{E} =0$, equation (\ref{e8}), for $M=m_e$ and $Q=e$, as required by the 
single-particle equivalence principle.

It is to be noted that the Witteborn-Fairbank experimental observations on the free fall of electrons have also been confirmed by Jain et \textit{al.} for electrons in superconductors subject to 
the earth gravitational field \cite{Jain}. Therefore the single-particle equivalence principle can also account for the behavior of Cooper pairs in homogeneous gravitational fields, thus 
confirming its applicability to quantum systems. 

\subsection{\label{section_2_2}Electric Charges with Constant Acceleration do not Radiate}

According to Einstein's principle of equivalence an electric charge in free fall, under the single influence of a homogeneous gravitational field, is locally indistinguishable from an inertial 
frame (which by definition is not accelerated), therefore it cannot radiate electromagnetic waves. If this would not be the case we could discriminate an inertial frame from a frame in free fall 
in a constant gravitational field, by simply monitoring if an electric charge attached to the frame radiates or not. In other words a charge at rest in a homogeneous gravitational field should 
radiate, if it does when attached to a uniformly accelerating reference frame \cite{Singal}. The single-particle principle of equivalence requires that gravitational acceleration and acceleration 
caused by a constant electric field, are both indistinguishable from inertial acceleration (caused only by the relative acceleration of a reference frame away from any fields), 
$a=qE/m_i= m_p \, g/m_i$. Therefore if an electric charge cannot radiate in a constant gravitational field, it should not be able to radiate neither when accelerated only by a constant homogeneous 
electric field. What does experiment say?

In electrodynamics it turns out that the power radiated by an accelerated charge is deduced from the work done against the force of radiation resistance \cite{Feynman} \cite{Lorrain}, it can be 
written
\begin{equation}
\frac{dW}{dt}=-k\frac{2}{3} \frac{e^2}{c^3} \,\, \vec{v} \,\, \cdot \dot{\vec{a}}, \label{e9}
\end{equation}
where $\vec{v}$ and $\dot{\vec{a}}$ are the speed and the time varying acceleration of the electric charge (to which the radiation reaction force is proportional), and $k=1/4\pi\epsilon_0$ is 
Coulomb constant with $\epsilon_0$ the vacuum permittivity. From equation (\ref{e9}) it is clear that electric charges do not radiate when accelerated in a uniform electric field. An electric 
charge only radiates when its acceleration vector is not constant. Like for example when an electric charge experiences a centripetal acceleration. The acceleration in the Larmor formula, 
$\frac{dW}{dt}=-\frac{2}{3} k \frac{e^2}{c^3}a^2$, is a centripetal acceleration, therefore it is not a constant acceleration (its direction changes with time along the particle's trajectory). 
Thus an electric charge moving along a circle does radiate electromagnetic energy, but an electric charge moving along a straight line under the single influence of a constant acceleration does 
not radiate (because the acceleration has constant direction, and constant magnitude) . The acceleration of the electric charge, as a vector quantity (direction, magnitude), must change over time 
in order to radiate electromagnetic energy.

Therefore in a uniform electric field, an electric charge behaves exactly like in a uniform gravitational field, i.e it does not radiate! Since this is exactly what requires the single-particle 
equivalence principle, we conclude that the observed behavior of electrical particles in homogeneous gravitational and electric fields supports this principle.

It is to be noted that the Witteborn-Fairbank experiment, discussed in section \ref{section_2_1}, also demonstrated the absence of electromagnetic radiation from electric charges with constant 
acceleration, but for a completely different reason: It reported that electric charges do not fall in a constant gravitational field therefore they cannot radiate electromagnetic waves. Since a 
weak uniform electric field was applied through the entire drift tube, one could have also expected that any electromagnetic radiative effects would have affected the measured electrons' time of 
flight during the free fall, which was caused only by the acceleration communicated by the applied constant electric field (since the effective gravitational acceleration was observed to be null). 
However this could not be detected with Witteborn and Fairbank experimental design, since the predicted radiative energy gradient effects, (if present), would have been of the order of 
$10^{-30} eV/m$, well below the sensitivity threshold of the experiment, which was $10^{-11} eV/m$. It would be worth revisiting Witteborn and Fairbank experiment to test this effect in particular 
\cite{Dittus}.

In any case we reach the conclusion that the radiative behavior of the electron in Witteborn-Fairbank experiment is consistent with the fundamental explanation of this experiment in terms of the 
single-particle equivalence principle, as discussed in section \ref{section_2_1}.

\subsection{\label{LEH} London Moment in Rotating Superconductors and the Single Particle Equivalence Principle in Superconductors}
In the absence of any external magnetic field, when a superconductor is set into rotation it generates a homogeneous magnetic field (which is also present in the bulk of the superconductor), hence 
acquiring a magnetic moment called the London moment \cite{Becker}. This is commonly accounted for by the quantum properties of superconductors. The electrons in the superconductor which are 
responsible for the London moment, are pairs of electrons which behave as a single particle called Cooper pairs (with twice the electron mass and charge). The supercurrent of Cooper pairs flows 
without friction through the superconductor's crystalline lattice, and forms a Bose Einstein condensate, described by one single wave function:
\begin{equation}
\psi=\rho^{1/2} e^{i\theta} \label{l1},
\end{equation}
where $\rho$ is the electric density of the Cooper pair condensate, and $\theta$ is the phase of the wave function.

The canonical linear momentum of Cooper pairs, 
\begin{equation}
\vec{\pi}=2m\vec{v}+2e\vec{A},\label{l2}
\end{equation}
where $\vec{v}$ is the velocity of the Cooper pairs, and $\vec{A}$ is the magnetic vector potential in which the Cooper pairs evolve, is proportional to the gradient of the phase of the Cooper 
pair condensate.
\begin{equation}
\vec{\pi}=\hbar \nabla(\theta). \label{l3}
\end{equation}
From this relation one deduces that the Curl of the Cooper pair momentum must be null.
\begin{equation}
\nabla \times \vec{\pi}=0. \label{l4}
\end{equation}
Substituting equation (\ref{l2}) in equation (\ref{l3}) one obtains the London moment of the rotating superconductor.
\begin{equation}
\vec{B}=-\frac{m}{e} 2\vec{\omega}, \label{l5}
\end{equation}
where $2\vec{\omega}=\nabla \times \vec{v}$ is the rotating angular frequency of the superconductor. It is remarkable that the generated homogeneous magnetic field is independent of any electric 
current, or material type, or material volume, and that it can be deduced directly from the single-equivalence principle in magnetic fields, cf section \ref{section_1_2_a}, which states that 
Coriolis and magnetic forces are indistinguishable for a set of charged particles with identical charge-to-mass ratio $e/m$:
\begin{equation}
2m\vec{v} \times \vec{\omega}=-e\vec{v} \times \vec{B}. \label{l6}
\end{equation}
From equation (\ref{l6}) it is straight forward to deduce the London moment, equation (\ref{l5}), and its reverse effect: The Einstein-de Haas effect \cite{Einstein} \cite{Frenkel}:
\begin{equation}
\vec{\omega}=-\frac{1}{2}\frac{e}{m}\vec{B}\label{l7}
\end{equation}
which has been observed in ferromagnetic materials when set into rotation when subject to a applied magnetic field $\vec{B}$.

More generally one observes that the single-particle equivalence principle is fully integrated in the generalized London equations describing the physical properties of superconductors in 
electromagnetic and weak gravitational, and gravitomagnetic fields obtained from the linearization of Einstein field equations \cite{Ross}. 

From this perspective the first London equation would result from the equivalence between inertial forces and electric and gravitational forces.
\begin{equation}
\vec{a}=-\frac{e}{m}\vec{E}-\vec{g}, \label{l8}
\end{equation}
where $\vec{a}$ is the acceleration of cooper pairs, $\vec{E}$ is the electric field, and $\vec{g}$ is the gravitational field. The second London equation would result from the equivalence between 
inertial Coriolis forces, in rotating systems, and magnetic and gravitomagnetic forces.
\begin{equation}
2\vec{\omega}=-\frac{e}{m} \vec{B}-\vec{B_g},\label{l9}
\end{equation}
where $\vec{\omega}$ is the angular velocity of Cooper pairs, $\vec{B}$ is the magnetic field $\vec{B_g}$ is the gravitomagnetic field, which has units of $rad.s^{-1}$ \cite{hobson}.

\section {\label{Conclu} Conclusions}
The present consensus in the scientific community is that the equivalence principle still needs to be tested for electrically charged matter \cite{Dittus}, and that free falling radiative 
charged particles in a gravitational field is still a conceptual challenge for the theory of general relativity, widely discussed by theoretical physicists\cite{Tamburini}. This seems to indicate 
that the current formulation of the equivalence principle at the basis of the theory of general relativity might have difficulties to make compatible electrically charged matter with a geometric 
description of the gravitational interaction.  

The extension of the principle of equivalence to electromagnetic fields, as introduced in \cite{Ozer_2000} and recapitulated in the present work, seems to be the easiest manner to unify gravitation
 with electromagnetism in a geometric framework, that can explain the observed null relative acceleration between charged particles in free fall in a constant gravitational field 
(Witteborn-Fairbank experiment), and the absence of electromagnetic radiation for freely falling electric charges, as well as the London moment in superconductors. The fact that the 
single-particle equivalence principle, formulated in section \ref{section_1_2}, is consistent with the predictions of quantum mechanics for what concerns the properties of superconductors 
(action of gravity on cooper pairs cf. section \ref{section_2_1}, and magnetic London moment cf. section \ref{LEH}), contributes to affirm its fundamental and universal character.

In essence, "the single-particle equivalence principle" reveals that inertia is the bridge between electromagnetism and gravitation if one postulates that \textit{inertia is relative not only to 
space-time geometry, but also to families of particles sharing the same charge-to-mass ratio, and the same gravitational to inertial mass ratio}. Thus one could account for electromagnetic forces 
in terms of inertial forces, with respect to sets of particles with identical $q/m$ ratio. Like for the case of gravitation, this allows to account for the fundamental physical nature of 
electromagnetism in terms of spacetime curvature, in line with equation (\ref{e5}). 

In table \ref{tab:inertia} we summarize the phenomenological framework supporting the proposed extension of the equivalence principle to electrically charged matter, as discussed in the present 
paper.

\begin{table}
\centering
\begin{tabular}{l|c|r}
Electromagnetism & Inertia & Gravitation \\\hline
$(q/m_i)E$ & $a$ & $(m_p/m_i) g$ \\
$(q/2m_i)B$ & $\Omega$ & $(m_p/2m_i) B_g$ 
\end{tabular}
\caption{\label{tab:inertia}Electromagnetism and gravitation are unified by inertia. A inertial frame is achieved when inertial acceleration $a$ and angular rotation of a rotating frame 
$\Omega$ vanish. Inertial acceleration is indistinguishable from gravitational, $(m_p/m_i) g$ (weak equivalence principle), and electric, $(q/m_i)E$, accelerations 
(supported by Witteborn-Fairbank experiments, and the absence of radiation from electric charges in free fall in gravitational fields). Angular rotation is indistinguishable from magnetic 
Larmor rotation frequency, $(q/2m_i)B$ (supported by the London moment in rotating superconductors, and the Einstein-de Haas effect), and from the Lens-Thirring angular rotation frequency, 
$(m_p/2m_i) B_g$ (which has been confirmed by Gravity Probe B measurements \cite{GPB}), where $B_g$ is the so called gravitomagnetic field with units of $rad/s$.}
\end{table}

The next steps would be to extend the theory of general relativity to electromagnetism on the basis of the "single-particle equivalence principle", which has already been done in 
\cite{Ozer_2000}, and to attempt the direct experimental detection of the various space-time curvature effects generated by electric charges \cite{Ozer_1999}, in analogy with similar effects 
caused by gravitational masses, like for example the:
\begin{enumerate}
\item Deflection of electrons by a charged sphere in a vacuum chamber,
\item Electric and magnetic deflection of light (similar to the gravitational deflection of light),
\item Generation of electrical geometry waves (similar to gravitational geometry waves which are currently called also "gravitational waves") by rotating electric dipoles \cite{Clovis}.
\end{enumerate}


\begin{thebibliography}{9}
\bibitem{note_1} It would be worth reproducing Witteborn-Fairbank experiment using freely falling ions instead of electrons, in order to test further the interpretation of this experiment in 
terms of the single-particle equivalence principle.
\bibitem{Clovis} To be discussed in detail in a separate dedicated communication
\end{thebibliography}
\end{document}